\documentclass[aip,jcp,reprint]{revtex4-1}
   \usepackage{amsmath}   %
   \usepackage{bm}        
   \usepackage{amsfonts}  
   \usepackage{amssymb}   %
   \usepackage{graphicx}  
   \usepackage{epstopdf}  
   \usepackage{braket}
   \DeclareGraphicsExtensions{.eps}
   \usepackage[bookmarks=false]{hyperref} 
   \hypersetup{ 
    colorlinks=true,
    citecolor=blue,
    linkcolor=blue,
    urlcolor=blue}
   \usepackage{url} 
   \usepackage[all]{hypcap} 
   \usepackage{multirow} 

\renewcommand{\thispagestyle}[1]{}

\begin{document}
\title{Many-body dispersion effects in the binding of adsorbates on metal surfaces}
\author{Reinhard J. Maurer}
\affiliation{Department of Chemistry, Yale University, New Haven, Connecticut 06520, USA}
\author{Victor G. Ruiz}
\author{Alexandre Tkatchenko}
\affiliation{Fritz-Haber-Institut der Max-Planck-Gesellschaft, Faradayweg 4-6, D-14195 Berlin, Germany}

\begin{abstract}
A correct description of electronic exchange and correlation effects for molecules in contact with extended (metal) surfaces is a challenging task for first-principles modeling. In this work we demonstrate the importance of collective van der Waals dispersion effects beyond the pairwise approximation for organic--inorganic systems on the example of atoms, molecules, and nanostructures adsorbed on metals. We use the recently developed many-body dispersion (MBD) approach in the context of density-functional theory [Phys. Rev. Lett. \textbf{108}, 236402 (2012); J. Chem. Phys. \textbf{140}, 18A508 (2014)] and assess its ability to correctly describe the binding of adsorbates on metal surfaces. We briefly review the MBD method and highlight its similarities to quantum-chemical approaches to electron correlation in a quasiparticle picture. In particular, we study the binding properties of xenon, 3,4,9,10-perylene-tetracarboxylic acid (PTCDA), and a graphene sheet adsorbed on the Ag(111) surface. Accounting for MBD effects we are able to describe changes in the anisotropic polarizability tensor, improve the description of adsorbate vibrations, and correctly capture the adsorbate--surface interaction screening. Comparison to other methods and experiment  reveals that inclusion of MBD effects improves adsorption energies and geometries, by reducing the overbinding typically found in pairwise additive dispersion-correction approaches.
\end{abstract}

\maketitle 

\section{Introduction}
\label{intro}

Atoms, molecules, and extended nanostructures interacting with solid surfaces are omnipresent in modern materials~\cite{Kronik2010}. Obtaining a detailed understanding of the physical and chemical properties of metal-adsorbed molecules is of paramount importance in a wide variety of fields. However, the description of such complex systems poses a significant challenge to first-principles modeling by featuring all limiting cases of chemical bonding ranging from the delocalized nearly-free electrons in the metal to directional covalent bonds in the adsorbate. In addition to these limiting cases, the description of adsorbate-surface bonding and dispersion interactions create a range of additional challenges. Many recent works suggest that dispersion interactions play a dominant role in many realistic hybrid organic--inorganic materials \cite{Kim2006,Tkatchenko2010, Tkatchenko2012,DiStasio2014,Liu2014}. The interplay between localized and delocalized states at such an interface can be translated into the relevance of both typically distinguished domains of electron correlation: the ``static correlation'' or inherent state-degeneracy that underlies the substrate metallic states, as well as the ``dynamic correlation'' that governs the long-range dispersive interactions induced by the quantum-mechanical fluctuations within the combined adsorbate--substrate system. Quantitatively or often even qualitatively correct first-principles treatment will have to efficiently account for both of these effects \cite{Rohlfing2008,Schimka2010}. Whereas wavefunction-based approaches to electron correlation yield an excellent description of the latter \cite{Manby2006,Nolan2009,Booth2013}, density-functional theory (DFT), specifically in its semi-local approximations based on the homogeneous electron gas, is able to describe the metal electronic structure relatively well \cite{Jones1989, Marsman2008}. 

Metallic states can be treated correctly by utilizing semi-local approximations to DFT such as the local-density approximation (LDA) or the generalized gradient appoximations (GGAs) that satisfy the homogeneous electron gas limit \cite{Jones1989, Perdew1992}. On the other hand, long-range correlation and dispersion interactions of isolated systems is best treated with well-established quantum-chemical approaches, such as the coupled cluster (CC) method \cite{Bartlett2007}. However, long-range correlation effects in solids or surfaces such as dispersion interactions between adsorbate and substrate, are still out of reach for these approaches and completely amiss in semi-local DFT \cite{Jurecka2007,Klimes2012}. The resulting lack of van der Waals (vdW) interactions can lead to failure to find any stable adsorbate structures \cite{McNellis2009, Liu2012}. Accurate treatment of long-range correlation and their manifestation as dispersion interaction is key to correctly describe the bonding of adsorbate-surface complexes. While many approaches beyond semi-local DFT exist trying to incorporate an improved description of exchange and correlation, either on the basis of the adiabatic connection fluctuation-dissipation theorem (ACFDT) \cite{Olsen2013, Olsen2014} or many-body perturbation theory \cite{Aryasetiawan1998, Ren2011,Caruso2013}, it may still be desirable to retain the simplicity and computational efficiency of semi-local functionals, but somehow incorporate a physically correct description of long-range correlation as it is given by wavefunction approaches.

Empirical pairwise additive dispersion-correction schemes on top of DFT, such as DFT incl. dispersion (DFT-D) in the variants proposed by Grimme \cite{Grimme2006,Grimme2010} can be seen as a first step to form such a hybrid approach. A simple physical form of the leading-order dispersion term is included and the corresponding empirical atomic parameters such as dispersion coefficients $C_6$, atomic polarizabilities, and van der Waals radii are precalculated and tabulated. A transition to the pure DFT treatment is assured by damping the vdW interactions at close interatomic distances. A more general and less empirical treatment of dispersion interactions is provided by the approaches of Becke and Johnson \cite{Becke2005} or Tkatchenko and Scheffler (TS) \cite{Tkatchenko2009}, where the dispersion parameters are functionals of the electron density calculated from DFT. The corresponding parameters therefore adapt to the chemical environment of the atom. Due to their success and efficiency, pairwise additive treatments of dispersion are now standard practice in gas-phase quantum chemistry in describing molecular aggregates and also yield a qualitatively correct description of metal-adsorbed molecule geometry and energetics as has been shown in the case of benzene and azobenzene adsorbed on coinage-metal surfaces \cite{McNellis2009, Liu2012}. Nevertheless, comparison to experiment has shown that a simple pairwise treatment of dispersion can induce significant overbinding, hence quantitative or ``predictive-quality'' description cannot be achieved in general \cite{Mercurio2010}. This is due to the fact that many-body correlation effects and Coulomb-screening within the substrate play a significant role in these systems \cite{Ruiz2012}. Accounting for the latter by effective renormalization of metal $C_6$ coefficients, as done in the vdW$^{\mathrm{surf}}$ scheme,\cite{Ruiz2012} has improved adsorbate geometries significantly \cite{Mercurio2013}, albeit at a remaining overbinding in adsorption energies. Furthermore, most studies of molecules on surfaces have concentrated on the limit of low coverage. Obviously, many-body dispersion effects will become even more pronounced for dense molecular monolayers and multilayers adsorbed on surfaces.

A different approach to include dispersion interactions in DFT is given by the vdW-DF type of density functionals, where an additional non-local correlation contribution is added to a semi-local exchange-correlation functional \cite{Dion2004, Klimes2011}. The non-local correlation is described by a two-point integral over the electron density and a given integration kernel. Although the correlation integral is of a two-body form, higher-order semi-local contributions can be effectively incorporated in the formulation of the kernel. Recent improvements in computational efficiency \cite{Roman-Perez2009} and performance \cite{Klimes2010, Klimes2011} have triggered a more widespread use of vdW-DF specifically in the context of metal-surface adsorption \cite{Li2012}. Problems of inconsistent exchange treatment in earlier versions, resulting in systematic underbinding, are being addressed in more recent versions, such as the vdW-DF-cx functional \cite{Berland2014,Berland2014a}.

An efficient long-range approach to correlation that goes beyond pairwise dispersion has recently been proposed in the form of the many-body dispersion or MBD method \cite{Tkatchenko2012,MBD-JCP}. In the MBD approach, the problem of calculating the long-range correlation of a set of atoms in a molecular arrangement is recast to calculating the correlation energy of a coupled system of quantum harmonic oscillators (QHO) interacting via the long-range dipole potential. Both, many-body contributions and screening effects due to mutual polarization of the QHOs, are accounted for and a range-separation in terms of the Coulomb potential facilitates the connection to the DFT functional \cite{DiStasio2014}. This approach quantitatively describes interaction energies of a wide variety of systems including many well established gas-phase testcases \cite{DiStasio2012, Tkatchenko2012} as well as molecular crystals \cite{Reilly2013} and supramolecular arrangements \cite{Ambrosetti2014} at minimal overhead compared to the pure DFT calculations. Furthermore, the foundation of MBD on the ACFDT enables a systematic improvement in terms of better approximations to the response kernel or using well-established quantum-chemical techniques for electron correlation as applied to coupled oscillators. 

In this work we aim to assess and establish the MBD approach for hybrid organic--inorganic systems, specifically organic adsorbates on metal surfaces, a field in which MBD contributions to binding play a dominant role. After shortly revisiting the method and discussing specific issues in the context of metal surfaces, we investigate the relevance of MBD interactions for an atom, a molecule, and an extended nanostructure adsorbed on a metal Ag(111) surface. We find that in all cases MBD interactions are highly important in order to account for the correct interaction energy and even more so to correctly describe response properties such as vibrational frequencies or the polarizability of the system. Comparing to experiment and other simulation approaches, we find significant quantitative improvement by including many-body dispersion effects.  We conclude the work by shortly outlining the remaining challenges to establish the many-body dispersion approach as a contender in the modeling of molecule--surface systems.

\section{The MBD Method}
\label{methods}

In the following we will review the physical foundations of the MBD method as it was recently published and analyzed in detail elsewhere~\cite{Tkatchenko2012, DiStasio2014, MBD-JCP}.
The quantum-mechanical electron correlation energy, which contains the dispersion energy of a system, can be calculated from the microscopic density-density response function $\chi(\mathbf{r},\mathbf{r}',i\omega)$ using the ACFDT \cite{Bohm1953,Gell-Mann1957,DiStasio2014}:
\begin{align}\label{eq-ACFDT}
 E_c &= -\frac{1}{2\pi}\int_0^{\infty} d\omega \\ \nonumber
 & \times\int_0^1 d\lambda \mathbf{Tr}[(\chi_\lambda(\mathbf{r},\mathbf{r}',i\omega)-\chi_0(\mathbf{r},\mathbf{r}',i\omega))v(\mathbf{r},\mathbf{r}')] .
\end{align}
In this formalism the response function $\chi$ at a certain interaction strength $\lambda$ is calculated self-consistently from a non-interacting reference response function $\chi_0(\mathbf{r},\mathbf{r}',i\omega)$ using a Dyson-like screening equation \cite{Lu2010}:
\begin{equation}\label{eq-Dyson}
 \chi_{\lambda} = \chi_0 +\chi_0(\lambda v+f_{xc})\chi_{\lambda},
\end{equation}
with $v$ being the Coulomb potential and $f_{xc}$ being the exchange-correlation kernel. Approximations to the above equations typically vary in the initial guess for the non-interacting response function $\chi_0$ and in the approximations to $f_{xc}$ (e.g. $f_{xc}=0$ in the random-phase approximation (RPA)). Typically $\chi_0$ is constructed using a set of effectively independent particles or quasiparticles \cite{Adler1962,Wiser1963}, such as Hartree-Fock (HF) states or Kohn-Sham (KS) states as a result of Hartree-Fock or DFT calculations \cite{Hohenberg1964, Kohn1965}. These (quasi)-independent states are renormalized with respect to the Coulomb interaction via Eq. (\ref{eq-Dyson}). Approaches of this kind to calculate the correlation energy have recently been established using the RPA on top of single-particle states from DFT \cite{Ren2012a, Olsen2013}. Even in the case of RPA, solving the above equations quickly becomes computationally intractable with system sizes that are typically needed to model realistic materials.

The main idea behind the DFT+MBD approach is to find an alternative efficient way to construct $\chi_0$ and solve the correlation problem with the help of an intermediate set of quasiparticle states. The effects of mutual instantaneous polarization and depolarization of (valence) electrons can be recast into a system of effective quantum harmonic oscillators (QHOs) or Drude quasiparticles \cite{Jones2013}. The electron density around every atomic nucleus is represented by such a three-dimensional QHO with an effective width, mass, and frequency that connect to the polarizability and dipolar response of the valence electrons. This so-called coupled fluctuating dipole model has proven very successful in describing long-range interaction and polarization \cite{Donchev2006,Cole2009, Berthoumieux2010}. In this picture the non-interacting response $\chi_0$ of a system simplifies to a simple product of individual localized QHO quasiparticle response functions. The corresponding interacting response function and correlation energy can then efficiently be calculated within the framework of Eqs.(\ref{eq-ACFDT}) and (\ref{eq-Dyson}) using any approximate quantum chemistry method.

The challenge of recasting the long-range correlation problem to a set of QHOs lies in finding a connection to current electronic structure methods correctly treating short-range interactions, such as semi-local density-functional approximations. Treating the long-range correlation problem with QHO quasiparticles, interactions between these should only include terms not yet treated in the short-range via DFT. Correspondingly the quasiparticles need to already effectively contain short-range polarization and screening effects that are included in the DFT description. One therefore defines the frequency-dependent dipole polarizability of every QHO
\begin{equation}\label{eq-alphaTS}
 \alpha_p(i\omega)^{\mathrm{TS}} = \frac{\alpha_p[n(\mathbf{r})]}{1+(\omega/\omega_p[n(\mathbf{r}])^2} \quad ,
\end{equation}
via the static atomic polarizability $\alpha_p[n(\mathbf{r})]$ and the characteristic excitation frequency $\omega_p[n(\mathbf{r})]$ of the atom it models. These parameters can be extracted from the electron density as predicted by DFT using the Tkatchenko-Scheffler (TS) scheme \cite{Tkatchenko2009}. Therefore $\alpha_p(i\omega)^{\mathrm{TS}}$ includes hybridization effects due to the local environment as well as short-range exchange-correlation effects. Short-range interaction screening and anisotropic polarizability changes are furthermore included by renormalizing the polarizability with respect to the short-range Coulomb interaction of two spherical gaussian charge distributions associated with each pair of oscillators \cite{DiStasio2014}. The resulting QHOs (described by screened polarizabilities $\bar{\alpha}_p$ and screened characteristic frequencies $\bar{\omega}_p$) are quasiparticles that implicitly contain the short-range polarization due to the presence of other QHOs. The partition into long-range and short-range interactions is made using a standard range-separation technique, which, due to the arbitrariness of this partition, introduces a single range-separation parameter $\beta$, which has to be predetermined once for a given exchange-correlation functional by adjustment to a dataset of accurate binding energies \cite{MBD-JCP}. The corresponding parameter is virtually independent of the employed reference dataset and hence only depends on the employed exchange-correlation functional.

The independent QHO states can be coupled to the long-range part of the Coulomb interaction by solving Eqs.~(\ref{eq-ACFDT}) and (\ref{eq-Dyson}) or by explicitly solving the corresponding quasiparticle Hamiltonian \cite{DiStasio2014, Jones2013}. In the current formulation of MBD the interaction between QHOs is only accounted for via dipole-dipole interactions, enabling an analytically exact and efficient solution at vanishing computational expense when compared to DFT. The eigenstates of this QHO Hamiltonian correspond to collective polarization states of the many-body system and the corresponding correlation energy is given as the difference between the zero-point energy of interacting and non-interacting QHOs. Therefore, although the initial non-interacting response $\chi_0$ is strictly local, the resulting interacting response is fully delocalized as would be the case in typical ACFDT-based approaches \cite{Olsen2013, Olsen2014}. In fact, solving the QHO quasiparticle Hamiltonian in the dipole approximation is equivalent to calculating the ACFDT correlation energy in the RPA for the same set of QHOs \cite{Tkatchenko2013}.

The DFT+MBD approach has been implemented in different software packages including FHI-AIMS~\cite{Blum2009}, VASP~\cite{VASP}, QuantumEspresso\cite{QuantumEspresso}, CASTEP~\cite{Clark2005,Reilly2013a}, ADF~\cite{ADF} and furthermore is available as an independent code version~\cite{MBD-code}. Its success has recently been shown for a variety of systems. For small to medium-sized molecules the MBD approach correctly captures anisotropy effects in the $C_6$ coefficients and the molecular polarizability tensor. Furthermore comparing to experimental dipole-oscillator strength distributions this amounts at an accuracy for effective $C_6$ coefficients in DFT+MBD of about 6.3\%, being almost equal to the high accuracy achieved in the TS method (5.5\%) \cite{DiStasio2014}. At the same time the binding energies are within 5\% mean absolute relative error (MARE) when compared to basis-set limit coupled-cluster singles, doubles and perturbative triples (CCSD(T)) calculations on the S22 data set of small intermolecular complexes \cite{Tkatchenko2012}. In both cases DFT+MBD significantly improves on pairwise-additive dispersion approaches. The strengths of DFT+MBD become evident for extended systems such as large supramolecular complexes \cite{Tkatchenko2012a, Ambrosetti2014} and molecular crystals, where many-body interactions play a paramount role. In the latter case, pairwise additive approaches fail to achieve the same accuracy they reach for gas-phase intermolecular interactions and overestimate the lattice energy. PBE0+MBD describes the lattice energy of 16 representative molecular crystals within a MARE of 4.5\% (PBE0+TS: 12.9\% MARE) when compared to lattice energies extrapolated from experimental enthalpies of sublimation \cite{Reilly2013}. 

Nevertheless, the dipole approximation utilized in the MBD method can sometimes turn out to be insufficient and systematic ways exist to extend this approach beyond the current state-of-the-art. For example, this can be achieved by solving the coupled QHO system with an attenuated Coulomb potential. This could be done approximately by employing well-established correlation techniques from wavefunction theory. In the current work we apply the DFT+MBD approach for metal-adsorbed organic molecules. In the context of dispersion interactions, these are especially challenging systems. On the one hand, the localized states of the adsorbate exhibit strong attractive van der Waals interactions, on the other hand the vanishing band gap of the metal substrate leads to a fully non-local collective substrate response that effectively screens the interactions, thereby reducing $C_6$ coefficients, van der Waals radii, and the corresponding polarizability \cite{Zaremba1976, Ruiz2012}. Although this effect can to some extent be accounted for by effective renormalization of pairwise parameters, as has been done in the DFT+vdW$^{\mathrm{surf}}$ method \cite{Ruiz2012}, additional many-body contributions can be expected to play a significant role. Although the quantum harmonic oscillators in the MBD calculation have a non-vanishing excitation gap and are initially localized, the interaction-induced delocalization of the polarizability is significantly closer to the correct metallic response when compared to a pairwise response. Furthermore, the dispersion energy results from an integration over all frequencies from 0 to $\infty$ [see Eq.~(\ref{eq-ACFDT})]. To supply a good starting point for the MBD scheme and to better capture the response of the extended substrate, the renormalized ``atom-in-a-bulk'' parameters used in this work are derived using the DFT+vdW$^{\mathrm{surf}}$ method and subsequently employed to parametrize the initial QHO response in the MBD scheme. Due to these reasons, we expect the MBD method to capture many-body correlation effects with reasonable accuracy even in metallic systems. Our expectation is fully confirmed by the performance of the DFT+MBD method for realistic molecule--surface systems discussed in the next section. 

\section{Results and Discussion}
\label{results}

In the following we will study the performance of the DFT+MBD approach for properties of atoms, molecules, and extended nanostructures adsorbed on metal surfaces. We study three representative adsorbate-substrate complexes, one of which is dominantly dispersion bound (Xe on Ag(111)), one large molecule adsorbed \emph{via} both covalent and dispersion interactions (PTCDA on Ag(111)), and an extended organic--inorganic interface (a graphene sheet on Ag(111)).

All calculations have been performed with the DFT+vdW$^{\mathrm{surf}}$ and DFT+MBD implementations in the all-electron full-potential FHI-AIMS code using numerical atomic-orbital basis sets \cite{Blum2009}. Throughout this work we employ the Perdew-Burke-Ernzerhof\cite{Perdew1996} (PBE) and Heyd-Scuseria-Ernzerhof\cite{Heyd2003} (HSE) functionals with dimensionless MBD range-separation parameters of $\beta=0.83$ and $\beta=0.85$ as well as tight numerical basis settings. The $\omega$ range-separation parameter in the HSE functional was chosen as 0.11 bohr$^{-1}$. The binding energy curve for Xe on Ag(111) was performed using the experimentally reported $ (\surd 3 \times \surd 3)\mathrm{R30^o}$ coverage structure for Xe residing at an on-top site of a 6-layered Ag(111) slab. We used a Monkhorst-Pack grid~\cite{Monkhorst:PackPRB1976} of $15 \times 15 \times 1 $ k-points in the reciprocal space and a vacuum gap of 20~\AA. The binding energy curve for PTCDA on Ag(111) was performed using a $\bigl(\begin{smallmatrix} 6 & 1 \\ -3 & 5 \end{smallmatrix} \bigr)$ surface unit cell in accordance to experimental results~\cite{Gloeckler:Seidel:etalSS1998}. For reasons of computational tractability, the PBE calculations have been performed with a 3-layered  Ag(111) metal slab generated with the PBE bulk lattice constant, a vacuum gap of 50~\AA, and a Monkhorst-Pack grid of $4\times4\times1$ k-points in the reciprocal space. For the vdW$^{\mathrm{surf}}$ and MBD calculations on top of PBE, we have used a 5-layered Ag(111) metal slab to converge the dispersion binding energy. Previous works using DFT+vdW$^{\mathrm{surf}}$ have shown that geometry relaxations induce relatively small deformations for the systems we study. In the case of PTCDA on Ag(111) the change in height of the terminal oxygen atoms corresponds to 0.09~\AA~\cite{Ruiz2012}. We therefore assume that this also holds for the MBD case. The geometry of graphene has been modeled using a 6-layered slab generated with the PBE bulk lattice constant of 4.14~\AA. It has been optimized using the respective dispersion correction method with the bottom-most layer frozen. In the case of DFT+MBD, numerical nuclear forces have been calculated using a finite-difference approach. 

\subsection{Xe adsorbed on Ag(111)}
\label{results-xe}

Noble gas adsorption on metal surfaces has been studied for a long time and acts as a prototypical example for dispersion interactions beyond the simple pairwise picture. As shown by Zaremba and Kohn \cite{Zaremba1976}, the collective response of the surface modifies the dispersion and terms beyond the leading $r^{-6}$ dependence become relevant. Taking into account this effect, the vdW$^{\mathrm{surf}}$ method projects this interaction into an effective pairwise treatment by renormalization of the $C_6$ coefficients, atomic polarizabilities, and vdW radii with respect to the dielectric response of the substrate, while however, still neglecting non-local effects beyond the pairwise approximation in the treatment of the combined adsorption system. 

\begin{figure}
 \centering\includegraphics[width=3.37in]{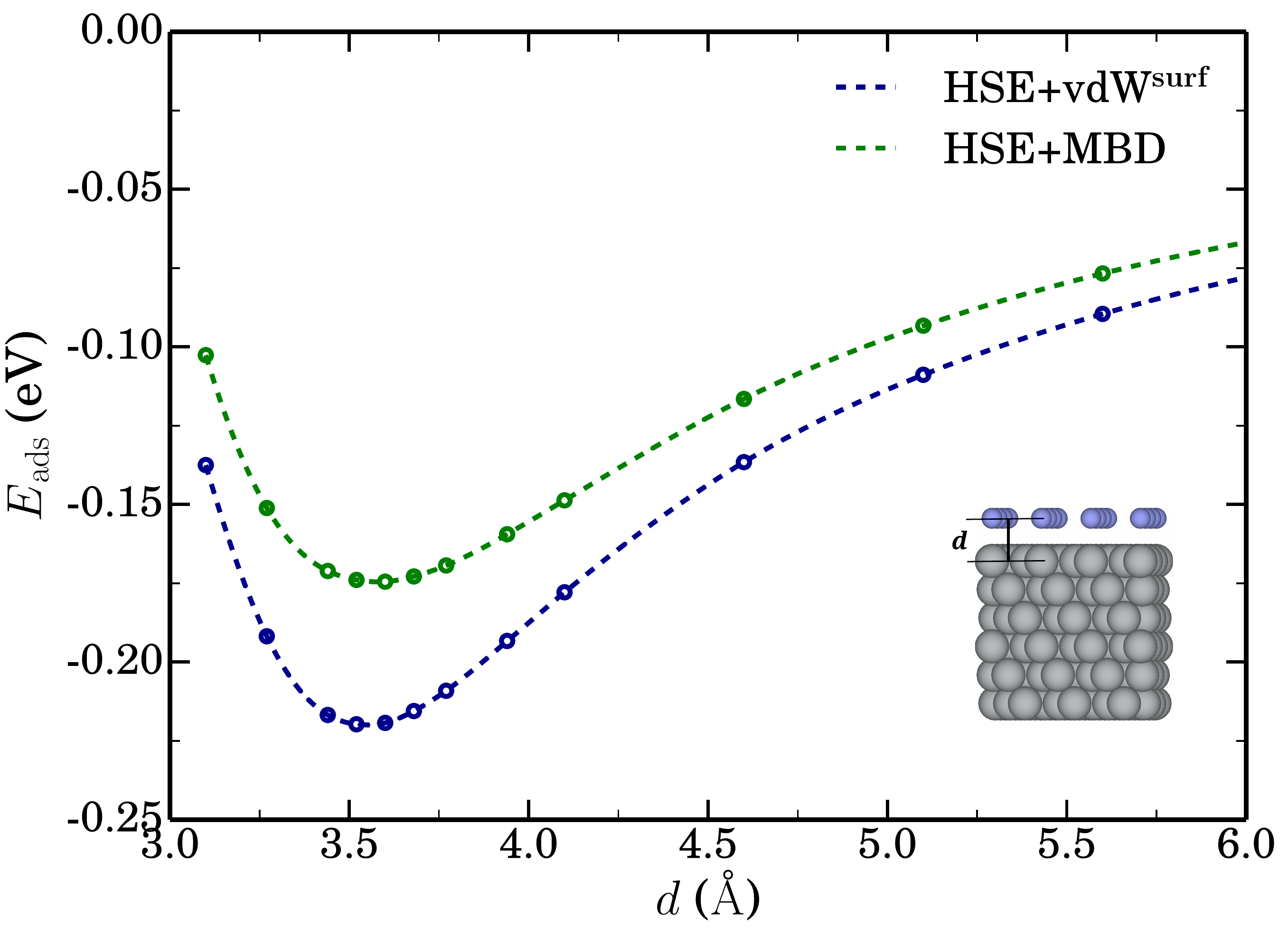}
 \caption{\label{fig-xe} Binding energy curve showing the adsorption energy $E_{\mathrm{ads}}$ as a function of vertical distance $ d $ for Xe on Ag(111) calculated with the HSE+vdW$^{\mathrm{surf}}$ (blue line) and HSE+MBD (green line) methods. The distance $ d $ is evaluated with respect to the position of the unrelaxed topmost metal layer. The results from the binding energy curve and their comparison to experiment can be found in \autoref{tab-xe}.}
\end{figure}

We have calculated the binding energy curve for Xe on Ag(111) including dispersion interactions with both the DFT+vdW$^{\mathrm{surf}}$ and DFT+MBD methods. The adsorption energy per adsorbed atom was calculated using 
\begin{equation}\label{Eq_EadsXeMe}
E_{\mathrm{ads}} = E_{\mathrm{Xe/Ag(111)}} - \left(E_{\mathrm{Ag(111)}} + E_{\mathrm{Xe}}\right), 
\end{equation}
where $ E_{\mathrm{Xe/Ag(111)}} $ is the total energy of the system (Xe monolayer + metal surface), $E_{\mathrm{Ag(111)}} $ is the energy of the bare slab, and $ E_{\mathrm{Xe}} $ is the energy of the isolated Xe gas atom, where all quantities correspond to the unrelaxed systems. The vertical distance $d$ was defined as the difference of  the position of the atom in the monolayer with respect to the position of the unrelaxed topmost metallic layer. \autoref{tab-xe} shows the optimal adsorption distance and energy using PBE and HSE as underlying exchange-correlation  functionals whereas \autoref{fig-xe} depicts the binding curve with the HSE+vdW$^{\mathrm{surf}}$ and HSE+MBD methods.

The adsorption height of Xe on Ag(111) has been studied experimentally using low-energy electron diffraction (LEED) \cite{Stoner1978} and synchrotron x-ray scattering \cite{Dai1999}, both assuming bulk truncation of the metal surface. Adsorption energetics have been studied using temperature-programmed desorption (TPD) \cite{McElhiney, Behm1986} and data from inelastic helium scattering \cite{Gibson1988}. Including error bars, the experimentally found adsorption energies and heights range from 0.18 to 0.23~eV and 3.45 to 3.68~\AA. The best estimates, as given by Vidali \emph{et al.}\cite{Vidali:Ihm:Kim:etalSSRep.1991}, are 3.6 $\pm$ 0.05 \AA\ for the adsorption height and an adsorption energy between 0.20 and 0.23 eV. For comparison, we also show the results coming from experiments  in \autoref{tab-xe}.

\begin{table}
\caption{\label{tab-xe}  Adsorption energy, vertical height, and perpendicular vibrational frequency of Xe adsorbed to Ag(111). Energy and  vertical height are given for the optimal value as extracted from the binding energy curve. For comparison, we also show results from experiments. The best estimates are, as given by Vidali \emph{et al.}\cite{Vidali:Ihm:Kim:etalSSRep.1991}, 3.6 $\pm$ 0.05 \AA\ for the adsorption height and an adsorption energy between 0.20 and 0.23 eV.}
 \begin{tabular}{ccccc} \hline
Xe/Ag(111)	&  $-E_{\mathrm{ads}}$ [eV] & $d$ [\AA] & $E_{\mathrm{vib}}$ [meV]  \\ \hline
  PBE+vdW$^{\mathrm{surf}}$	& 0.22 &  3.56 & 3.8  \\ 
  PBE+MBD	&  0.17  &  3.64  & 3.0 \\ 
   HSE+vdW$^{\mathrm{surf}}$	&  0.22  & 3.52 & 4.0 \\ 
  HSE+MBD	&  0.17  &  3.57  &  3.3 \\ \hline
  cRPA+EXX \cite{Rohlfing2008}	&  0.14  &  3.6   & 	 & 	\\ 
  \hline
  Exp.\cite{Gibson1988,McElhiney,Behm1986,Dai1999,Stoner1978}
				&  0.18 - 0.23  &  3.45 - 3.68  & 2.8  \\ \hline  
 \end{tabular} 
\end{table}

The results in \autoref{tab-xe} show an adsorption height of 3.56~\AA\ and an adsorption energy of 0.22~eV with the PBE+vdW$^{\mathrm{surf}}$ method, demonstrating that already on the level of pairwise-additive dispersion including the collective substrate response, the agreement with experiment is excellent. Inclusion of explicit many-body effects using the above presented MBD scheme yields, as expected, only small changes (see \autoref{tab-xe}). The vertical adsorption height is increased above 3.6~\AA\ and the adsorption energy is slightly reduced, suggesting a small repulsive contribution from higher-order terms beyond the leading $r^{-6}$ interatomic behavior. The energy dependence with respect to adsorption height suggests a very slow decay with distance from the surface for both methods.

It is clear that a balanced description in terms of accuracy between exchange and correlation can be an essential factor in determining the structural and energetic features in adsorption phenomena.  Having in mind this fact,  we have also calculated the binding energy curves using HSE as the underlying functional, thereby improving upon the description of electronic exchange effects. These curves are depicted in \autoref{fig-xe} and the results of these are shown in \autoref{tab-xe}. The adsorption energy is not substantially modified when comparing between PBE and HSE results, which is not surprising if we consider that the attractive part of the interaction in this system will mainly be contained in the correlation energy. In this sense, the adsorption energy is more sensitive to the description of the dispersion interactions than to the description of exchange. On the other hand, the adsorption height seems to be more sensitive to the choice of the exchange-correlation functional. The vertical adsorption height is 3.52~\AA\ with the HSE+vdW$^{\mathrm{surf}}$ method and 3.57~\AA\ with the HSE+MBD method. Inclusion of many-body effects using the MBD scheme yields a slight increment of 0.05~\AA\ in the adsorption height when HSE is the underlying exchange-correlation functional, in contrast to the larger increment of 0.08~\AA\ observed with the PBE functional. These facts suggest that the inclusion of screened short-range exact exchange, as found in the HSE functional, provides a more balanced description when coupled to the MBD scheme as part of the correlation energy. Although the effects may seem to be small in the case of Xe on Ag(111), these may become increasingly important in the adsorption of organic adsorbates on metal surfaces, where we find a much more complex interplay of interactions. As a final remark, it is worth to mention that many-body effects persist for large distances (larger than approximately 5.0 \AA) as the comparison between binding curves with the vdW$^{\mathrm{surf}}$ and the MBD methods in \autoref{fig-xe} shows. The discrepancy  between the binding curves at an infinite distance is given, in accordance to the definition in Eq. (\ref{Eq_EadsXeMe}), by the different values for the formation energy of the monolayer calculated with the vdW$^{\mathrm{surf}}$ and the MBD methods. 

Rohlfing and Bredow have studied Xe on Ag(111) using a correlation treatment based on the RPA including exact exchange (EXX)\cite{Rohlfing2008}, which accounts for many-body effects by explicitly calculating the long-range correlation energy of the system. The corresponding adsorption energy lies about 30~meV above our PBE+MBD and HSE+MBD results, whereas the adsorption height is found to be in good agreement with both experiment and only 0.04~\AA\ below PBE+MBD and 0.03 \AA\ above our HSE+MBD results. This suggests that explicit account of many-body dispersion does in fact reduce the binding strength in comparison to a pairwise treatment. In the latter case of correlation based on the RPA in Ref. \cite{Rohlfing2008}, this effect might be overestimated due to neglect of the exchange-correlation kernel, the underlying plasmon-pole approximation, and the fact that the response function of the system is not fully coupled, being calculated separately for substrate and adsorbate.

We have also computed the perpendicular vibrational frequencies of Xe on the metal surfaces to probe the curvature of the potential energy curves around the minimum in each case. For this, following previous works~\cite{Bruch:Cole:ZarembaPhysicalAds1997,DaSilva:Stampfl:SchefflerPRB2005,Chen:Al-Saidi:JohnsonPRB2011,Chen:Al-Saidi:JohnsonJPCM2012}, we have modeled the gas-surface adsorption potential with the following function given by the sum of repulsive and attractive dispersion interactions
\begin{equation}\label{Eq_FitAdsPotential}
 E_{\mathrm{ads}}(d) = \alpha_1 e^{-\alpha_{2} d} - \dfrac{C_3}{(d-Z_0)^3} + E_{\rm{ML}},
\end{equation}  
where $E_{\mathrm{ads}}(d)$ is the adsorption potential between Xe and the metal substrate at a distance $d$ from the surface and 
$E_{\rm{ML}}$ is a constant that corresponds approximately to the formation energy of the Xe monolayer. We have determined the 
parameters $\alpha_1$, $\alpha_2$, $C_3$, $Z_0$, and $E_{\rm{ML}}$ by fitting Eq.~(\ref{Eq_FitAdsPotential}) to the binding energy curve calculated with each method. The vibrational energy $E_\mathrm{vib}$ is then calculated using $E_{\mathrm{vib}} = h\nu = h/(2\pi)\sqrt{k_e/m_{\mathrm{Xe}}}$ where $\nu$, $h$, and $m_{\mathrm{Xe}}$ are the vibrational frequency, Planck's constant, and the mass of a Xe atom, respectively. The force constant $k_e$ corresponds to the second derivative evaluated at the minimum of the potential given by Eq.~(\ref{Eq_FitAdsPotential}). Following this procedure, the results for $E_{\rm vib}$ are given in \autoref{tab-xe}. Considering the absolute magnitude of this vibration known from experiment (2.8~meV) all methods perform very well, with explicit many-body dispersion reducing the vibrational energy closer towards the experimental value in the cases of PBE+MBD and HSE+MBD. This MBD-induced reduction in frequency is also visible from the width and curvature of the HSE+MBD binding energy in \autoref{fig-xe} when compared to HSE+vdW$^{\mathrm{surf}}$. 

\begin{figure}
  \centering\includegraphics[width=3.37in]{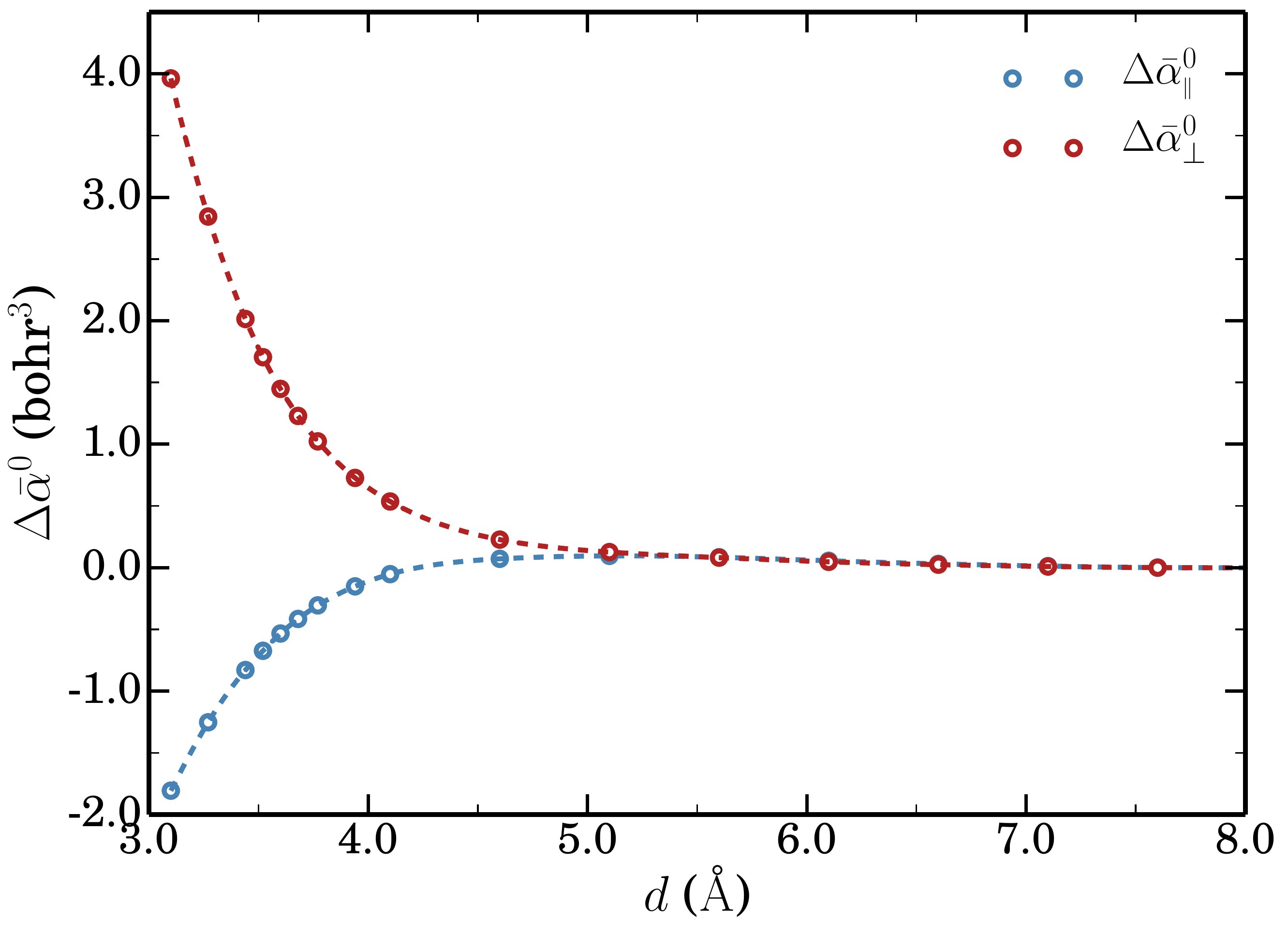}
  \caption{\label{fig-xe-alpha} Differential static polarizability $\Delta \bar{\alpha}^0$ of Xe on Ag(111) [see Eq. (\ref{Eq_delta alpha_XeMe})] as a function of the vertical distance $ d $ of the Xe monolayer as calculated with the HSE+MBD method.  Shown are the components parallel (blue curve) and perpendicular (red curve) to the surface plane.} 
\end{figure}

Although we are naturally interested in the computation of adsorption distances and energies and how they compare to experiments, we must also understand the improvements in the description of the physics and chemistry behind our methods. One of the novelties behind the MBD method is a more robust description for the polarizability of molecules and solids as the method includes many-body screening effects coming from the electrodynamic response of the system, a fact which leads to a more accurate description of dispersion interactions (see \autoref{methods}).
A more detailed understanding of these effects in the adsorption properties of Xe on Ag(111) can be gained by studying the changes in the static polarizability tensor of the system $\Delta \bar{\alpha}^0$ as a function of the adsorption height $d$ of the Xe monolayer. We have computed these changes using 
\begin{equation}\label{Eq_delta alpha_XeMe}
\Delta \bar{\alpha}^0_{(\raisebox{0.5pt}{$\scriptscriptstyle\parallel$}/\perp)} = \left. \bar{\alpha}^0_{(\raisebox{0.5pt}{$\scriptscriptstyle\parallel$}/\perp)} \right. _{\mathrm{Sys}} - \left( \left. \bar{\alpha}^0_{(\raisebox{0.5pt}{$\scriptscriptstyle\parallel$}/\perp)} \right. _{\mathrm{Ag(111)}} + \left. \bar{\alpha}^0_{(\raisebox{0.5pt}{$\scriptscriptstyle\parallel$}/\perp)} \right. _{\mathrm{Xe}} \right), 
\end{equation}
where $\bar{\alpha}^0_{\mathrm{Sys}} $ is the screened static polarizability of the complete system (Xe monolayer + metal surface), $\bar{\alpha}^0_{\mathrm{Ag(111)}}$ is the screened static polarizability of the bare slab, and $\bar{\alpha}^0_{\mathrm{Xe}}  $ is the screened static polarizability of the Xe monolayer in periodic boundary conditions. The treatment of the  dipole-dipole coupling between atoms in the MBD method naturally introduces anisotropy in the polarizability of the complete sytem \cite{DiStasio2014}. Due to the symmetry of the system, we observe two equivalent components in  the polarizability of the system which lie parallel to the plane of the surface; we denote their change as $\Delta \bar{\alpha}^0_{\raisebox{0.5pt}{$\scriptscriptstyle\parallel$}}$. The third component points in the direction perpendicular to the plane of the surface, we denote its change as $\Delta \bar{\alpha}^0_{\perp} $. \autoref{fig-xe-alpha} shows the results of Eq. (\ref{Eq_delta alpha_XeMe}) for each component as a function of the adsorption height $d$. The quantities shown in \autoref{fig-xe-alpha} correspond to the MBD calculations associated to each distance in the binding energy curve of \autoref{fig-xe}. 

The definition of $\Delta \bar{\alpha}^0$ in Eq. (\ref{Eq_delta alpha_XeMe}) lets us identify the distance upon adsorption at which the coupling of the components becomes relevant for the polarizability of the system. At large adsorption distances, the polarizability of the complete system, given by the first term in Eq. (\ref{Eq_delta alpha_XeMe}), is equal to the sum of its parts, found in the second term (in parenthesis) of  Eq. (\ref{Eq_delta alpha_XeMe}). This balance towards zero starts approximately at 5.5 \AA\ and becomes practically zero at distances greater than 7.5 \AA\ no matter which direction is considered. At distances lower than 4.5 \AA,  the coupling between Xe and Ag(111) starts to become relevant.  Upon reduction of the adsorption distance of the Xe monolayer, the polarization of the system in the direction parallel to the surface (given by the blue curve) is decreased in favor of an increasing polarization in the direction perpendicular to the surface (given by the red curve) due to the interaction with the surface. This is reflected in the negative values found in  $\Delta \bar{\alpha}^0_{\raisebox{0.5pt}{$\scriptscriptstyle\parallel$}}$ upon adsorption as well as the positive values in $\Delta \bar{\alpha}^0_{\perp} $. This behavior yields a stronger polarization of the system in the direction perpendicular to the surface plane at the equilibrium adsorption distance. Correspondingly, the increased polarizability towards the surface leads to larger interaction screening from many-body contributions, while at the same time dispersion interactions between Xe atoms are reduced. The changes in the anisotropic terms of the polarizability of the system may seem to be small in magnitude. However, these subtle changes induced by the coupling of the complete system could generate a preferential interaction along a specific direction parallel or perpendicular to the surface plane, yielding directionality in the formation of molecular monolayers. 


\subsection{PTCDA adsorbed on Ag(111)}
\label{results-ptcda}

\begin{figure}
 \centering\includegraphics[width=3.37in]{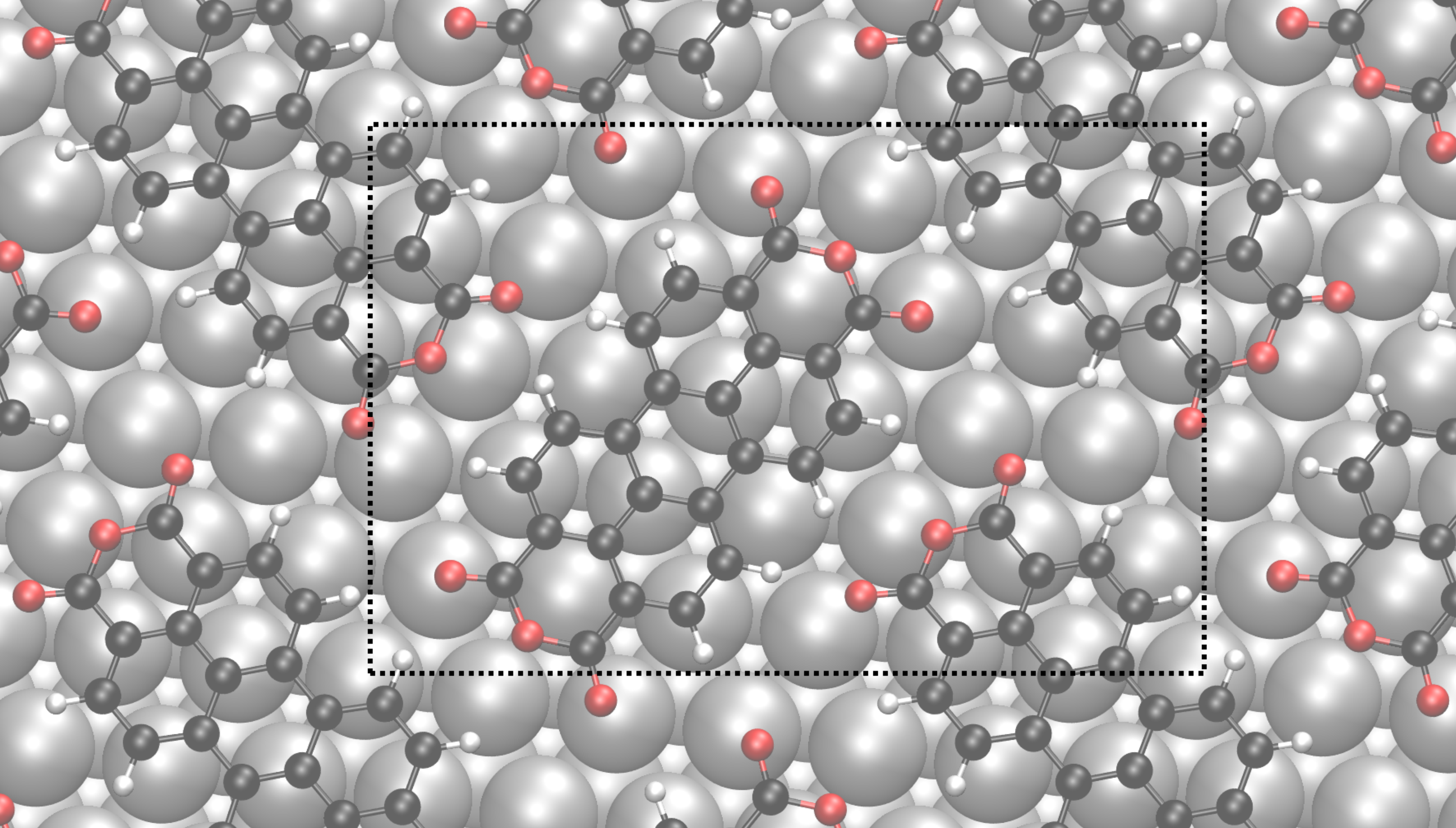}
 \caption{\label{fig-ptcda-model} PTCDA overlayer adsorbed to Ag(111) surface with the supercell geometry shown as a dotted line.}
\end{figure}

3,4,9,10-perylene-tetracarboxylic acid (PTCDA) is one of the best-studied large organic adsorbates on coinage metal surfaces, both experimentally \cite{Tautz2007,Hauschild2010, Mercurio2013b} and theoretically \cite{Vazquez2007, Rohlfing2008, Romaner2009, Ruiz2012}. PTCDA consists of two anhydride head-groups and an aromatic backbone. The molecule adsorbs in a densely-packed herringbone arrangement on the Ag(111) surface (see \autoref{fig-ptcda-model}), with the oxygens being tightly bound via covalent interactions and the aromatic backbone interacting with the surface via dispersion interactions \cite{Tautz2007}. Using current electronic structure methods, either based on pairwise dispersion-corrections or RPA-based approaches, the interaction energy as well as the vertical adsorption height of PTCDA adsorbed on the Ag(111) surface is on the upper range of what is expected to be the experimental adsorption energy (\emph{vide infra}). In the case of PBE+vdW$^{\mathrm{surf}}$, this happens even though the metallic substrate response has already been effectively accounted for in the dispersion parameters. As it has also been shown for the case of azobenzene adsorbed to Ag(111), neglecting this effect leads to even further overestimation of adsorption energies and adsorption heights \cite{Mercurio2013}. The remaining deviations to experiment, for example the overestimation of the adsorption interaction, may be ascribed to missing many-body dispersion contributions and the semi-local treatment of exchange interactions, both of which we will discuss below.

On the experimental side, the adsorption geometry and the individual atomic adsorption heights of PTCDA on Ag(111) are known from x-ray standing wave (XSW) measurements \cite{Hauschild2010}. The adsorption energy, typically extracted from TPD measurements, however, is not experimentally known due to the molecule being destroyed upon thermal desorption \cite{Zou2006}. However, Stahl \emph{et al.} have measured a binding energy of 1.16 $\pm$ 0.1~eV for  1,4,5,8-naphthalene-tetracarboxylic-dianhydride (NTCDA) on Ag(111) \cite{Stahl1998}, a smaller analogue of PTCDA containing the same number of terminal oxygen atoms but with a smaller aromatic backbone (40\% less molecule surface area). Based on surface-state photoemission data, Ziroff \emph{et al.} estimated the binding energy of PTCDA and its smaller analogue NTCDA on Au(111) by assuming purely physisorptive surface-binding and utilizing a connection to noble-gas adsorption on metals. The corresponding binding energies are reported as 2.0 and 1.5~eV for PTCDA and NTCDA on Au(111) \cite{Ziroff2009}. Contrary to that, Wagner \emph{et al.} report significantly higher binding energies of 2.6 and 1.7~eV for both molecules on Au(111) from force pulling experiments using an atomic force microscope (AFM)  \cite{Wagner2012}. Considering the TPD data on NTCDA on Ag(111) and the fact that two experiments place the additional binding energy of PTCDA on Au(111) compared to NTCDA on Au(111) to be within 33 to 66\%, we estimate the adsorption energy of PTCDA on Ag(111) to be between 1.4 and 2.1~eV (see \autoref{tab-ptcda}).

\begin{table}
\caption{\label{tab-ptcda} Adsorption energy and vertical height of PTCDA adsorbed to Ag(111). Energy and vertical height are given for the optimal value as extracted from the binding energy curve. For comparison, we also show results from experiments.}
 \begin{tabular}{ccc} \hline
PTCDA/Ag(111)	&  $-E_{\mathrm{ads}}$ [eV] & $d$ [\AA] \\ \hline
  PBE+vdW$^{\mathrm{surf}}$	& 2.50  &  2.89    \\ 
  PBE+MBD			&  1.77  &  2.94    \\ 
  vdW-DF$^{\mathrm{non-sc}}$ \cite{Romaner2009}  &  2.0  & 3.5 \\ 
  vdW-DF-cx\cite{Berland2014}	&  3.5 & 3.1  \\
  cRPA+EXX\cite{Rohlfing2008}	&  2.4 & 3.1  \\ 
  Exp. \cite{Hauschild2010, Stahl1998, Ziroff2009, Wagner2012}	
				&  (1.4 - 2.1)$^a$  &  2.86 $\pm$ 0.01   \\    \hline
 \end{tabular} 
 \begin{center} $^a$ estimate from experimental results: see text for more details \end{center}
\end{table}

\begin{figure}
 \centering\includegraphics[width=3.37in]{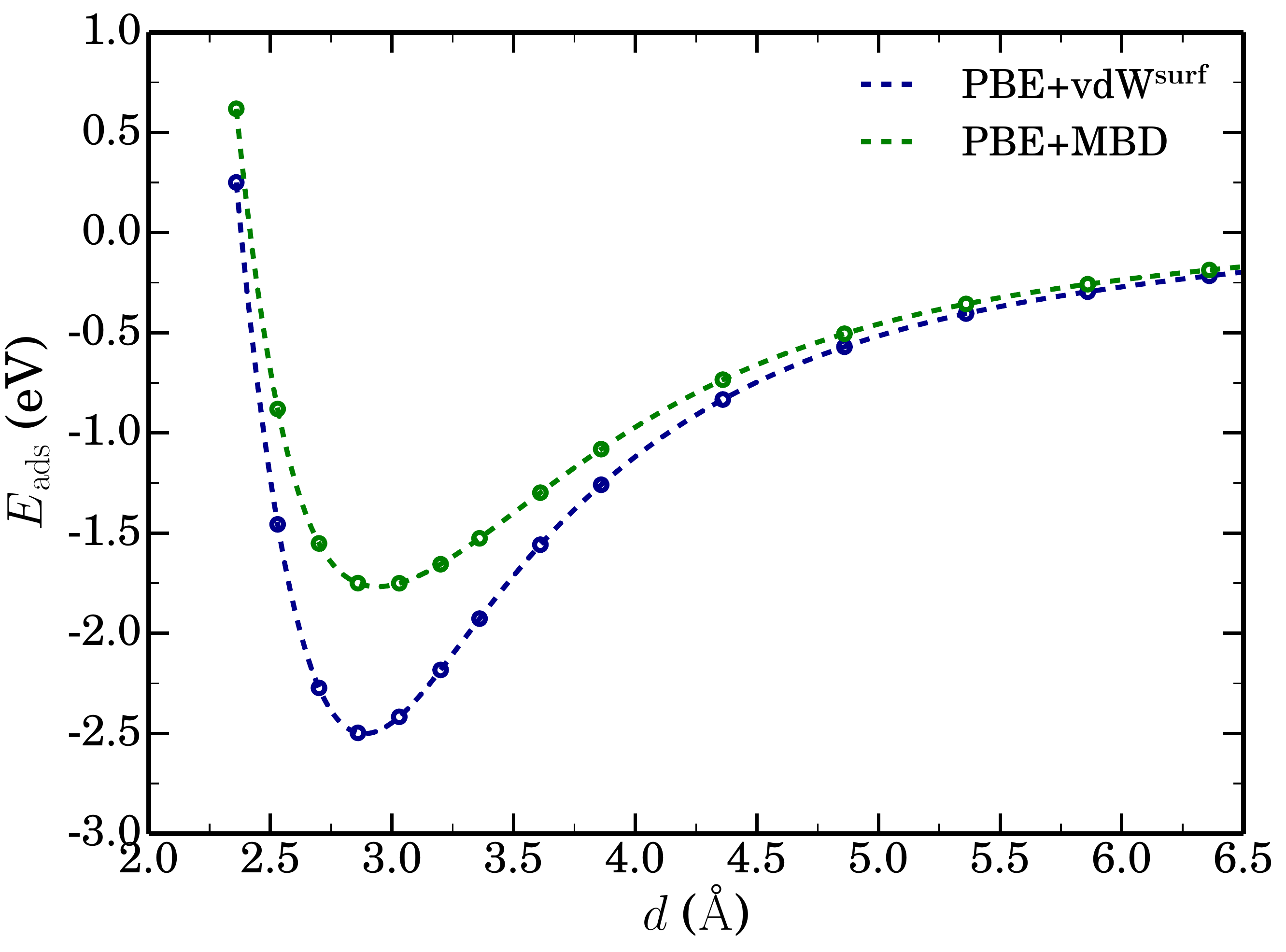}
 \caption{\label{fig-ptcda} Binding energy curve showing the adsorption energy $E_{\mathrm{ads}}$ as a function of vertical distance $ d $ for PTCDA on Ag(111) calculated with the PBE+vdW$^{\mathrm{surf}}$ (blue line) and PBE+MBD (green line) methods. The distance $ d $ is evaluated with respect to the position of the unrelaxed topmost metal layer. The results from the binding energy curve and their comparison to experiments can be found in \autoref{tab-ptcda}.}
\end{figure}

We have calculated the binding energy curve for a PTCDA monolayer on Ag(111) (two adsorbed molecules per unit cell, see \autoref{fig-ptcda-model})  including dispersion interactions with both the PBE+vdW$^{\mathrm{surf}}$ and PBE+MBD methods. The adsorption energy per molecule in the monolayer was calculated using 
\begin{equation}\label{Eq_EadsPTCDAAg}
E_{\mathrm{ads}} = \frac{1}{2} \left[ E_{\mathrm{PTCDA/Ag(111)}} - \left(E_{\mathrm{Ag(111)}} + E_{\mathrm{PTCDA}}\right) \right], 
\end{equation}
where $ E_{\mathrm{PTCDA/Ag(111)}} $ is the total energy of the system (PTCDA monolayer + metal surface), $E_{\mathrm{Ag(111)}} $ is the energy of the bare slab, and $ E_{\mathrm{PTCDA}} $ is the energy of the PTCDA monolayer in periodic boundary conditions, where all quantities correspond to the unrelaxed systems. The vertical distance $ d $ was defined as the difference of  the position of the monolayer with respect to the unrelaxed topmost metallic layer. \autoref{tab-ptcda} shows the optimal adsorption distance and energy using PBE as underlying exchange-correlation functional whereas \autoref{fig-ptcda} depicts the binding energy curve with both methods. Including many-body dispersion contributions via PBE+MBD, we find that the adsorption binding strength is reduced in comparison to PBE+vdW$^{\mathrm{surf}}$. This is reflected in  a reduced interaction energy and an increased adsorption height as depicted in \autoref{fig-ptcda}. Accounting for the higher-order correlation terms and the correct intermolecular polarization counteracts the dispersion energy of individual pairs of atoms stemming from the leading $r^{-6}$ term. From the binding-energy curves in \autoref{fig-ptcda} we further find that the curvature around the basin of attraction is reduced with the basin itself being widened. As in the case of Xe on Ag(111) this suggests both a reduced vibrational frequency and larger anharmonic contributions along the molecule-surface mode. The higher-order dispersion terms from the MBD come into effect at practically all distances considered in \autoref{fig-ptcda} reducing the dispersion energy. The MBD binding energy closely follows the vdW$^{\mathrm{surf}}$ binding energy only at adsorption distances larger than approximately 6.0 \AA. Upon further reduction of the adsorbate--surface distance, the electron density overlap is increased and the Pauli repulsion becomes the dominant term. At the same time, the many-body contributions become smaller by approaching a distance from the first metal layer of the order of the range-separation length.

By inclusion of explicit many-body dispersion we also find an improvement in performance when compared to experiment and literature data. Whereas PBE+vdW$^{\mathrm{surf}}$ is known to already yield a good description of adsorbate geometries \cite{Ruiz2012, Mercurio2013}, corresponding adsorption energies are systematically overestimated (see \autoref{tab-ptcda}). The PBE+MBD scheme yields an adsorption energy of 1.77~eV that lies within the estimated regime, and an adsorption height of 2.94 \AA\, which exceeds the experimental adsorption distance of 2.86~\AA\ by 0.08~\AA. Comparing to other calculations based on explicit correlation treatment, either on the RPA \cite{Rohlfing2008} or the vdW-DF level \cite{Romaner2009, Berland2014}, we find that PBE+MBD yields lower adsorption energies at smaller equilibrium distances from the surface. The overestimation of equilibrium adsorption distances is a well known issue for first and second generation vdW-functionals such as vdW-DF \cite{Klimes2011, Klimes2012}. However, this problem seems to have been remedied to some extent in the most recent vdW-DF-cx functional \cite{Berland2014}.
In the case of the PBE+MBD method, the adsorption distance is within 0.1 \AA\ of the value found in experiment.
Moreover, an accurate treatment of screened exchange in the underlying functional, via \emph{e.g.} HSE, will further improve the description of the geometry as happens in the case of Xe on Ag(111),  where we have observed a reduction of the adsorption distance by 0.07 \AA\ upon coupling the MBD energy to the HSE functional (see \autoref{results-xe}).
Finally, it is important to point out that our calculations for the binding-energy curve correspond to an unrelaxed system, that is a planar PTCDA monolayer adsorbed on an unrelaxed surface slab. Undertaking a full relaxation of the system using the MBD method will lead to 
the well known distortion of the molecule within the monolayer thereby yielding an improved description of the adsorption distance. 
For example, full geometry relaxation of PTCDA/Ag(111) using the PBE+vdW$^{\rm{surf}}$ method decreases the average adsorption distance by 0.09 {\AA} and increases the binding energy by 0.36~eV. The same change with PBE+MBD would still place the results within the expected experimental range.


\subsection{Graphene adsorbed on Ag(111)}
\label{results-graphene}

\begin{figure}
 \centering\includegraphics{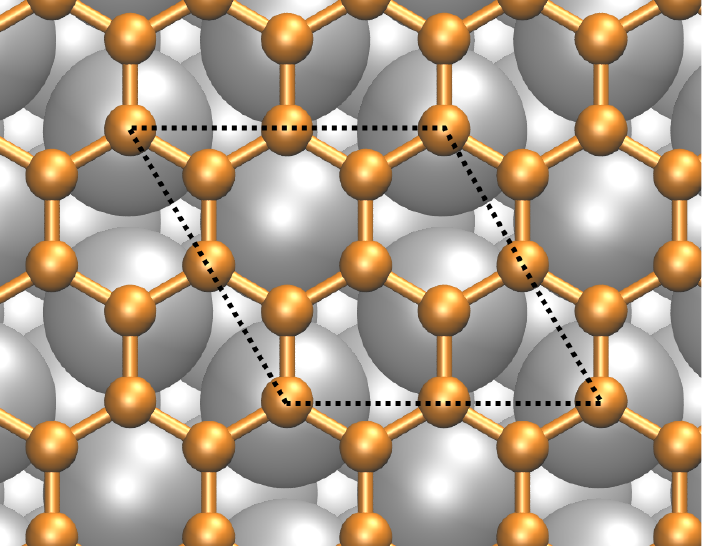}
 \caption{\label{fig-graphene} Graphene sheet adsorbed to Ag(111) in a $\sqrt{3}\times\sqrt{3}$ surface unit cell.}
\end{figure}

Graphene monolayers adsorbed on metal surfaces represent especially challenging systems with 
regards to dispersion interactions. Graphene sheets and more generally carbon nanostructures,
such as carbon nanotubes and fullerenes, exhibit different dispersion interactions than small 
organic molecules. This is apparent from the significant deviations of effective $C_6$ coefficients 
across different carbon-based materials  when accounting for polarization-induced screening \cite{Gobre2013}. 
The reason for this is a largely delocalized, collective polarization response to electric fields. 
Equally for a graphene monolayer interacting with an Ag(111) surface, screening effects and many-body dispersion contributions to the adsorption energy can be expected to be large. 
We have performed calculations for a graphene sheet commensurately adsorbed on a $\sqrt{3}\times\sqrt{3}$ Ag(111) slab (see Fig. \ref{fig-graphene}). In this adsorption geometry, the intramolecular carbon bonds in graphene are slightly elongated with 2.54~\AA, when compared to the 2.46~\AA\ of isolated graphene, however the unit cell is still small enough not to induce significant buckling in the graphene sheet due to the metal surface corrugation. We choose this unit cell for reasons of comparability to other studies \cite{Olsen2013, Loncaric2014} and computational tractability. We define the adsorption energy of graphene (Gr) per carbon atom as
\begin{equation}\label{Eq_EadsGrMe}
E_{\mathrm{ads}} = \left[E_{\mathrm{Gr/Ag(111)}} - \left(E_{\mathrm{Ag(111)}} + E_{\mathrm{Gr}}\right)\right] / N_C. 
\end{equation}

The PBE+MBD approach reduces the adsorption energy per atom of graphene on Ag(111) over 38\% compared to the pairwise PBE+vdW$^{\mathrm{surf}}$ approach. The resulting 45~meV per carbon atom are found at an equilibrium adsorption height of 3.23~\AA. As found for Xe and PTCDA on Ag(111), the equilibrium adsorption height is increased by inclusion of many-body effects. However the effect on the adsorption height can be considered strong when compared to Xe and PTCDA and stems from the large magnitude of many-body effects between metal and graphene that effectively screen the dispersion interactions. The resulting binding energy of 45~meV per carbon atom is close to the interlayer binding energy of crystalline graphite as predicted by PBE+MBD (48~meV/C atom) \cite{MBD-JCP}, suggesting a consistent description of many-body effects for nanostructured carbon.

\begin{figure*}
 \centering\includegraphics{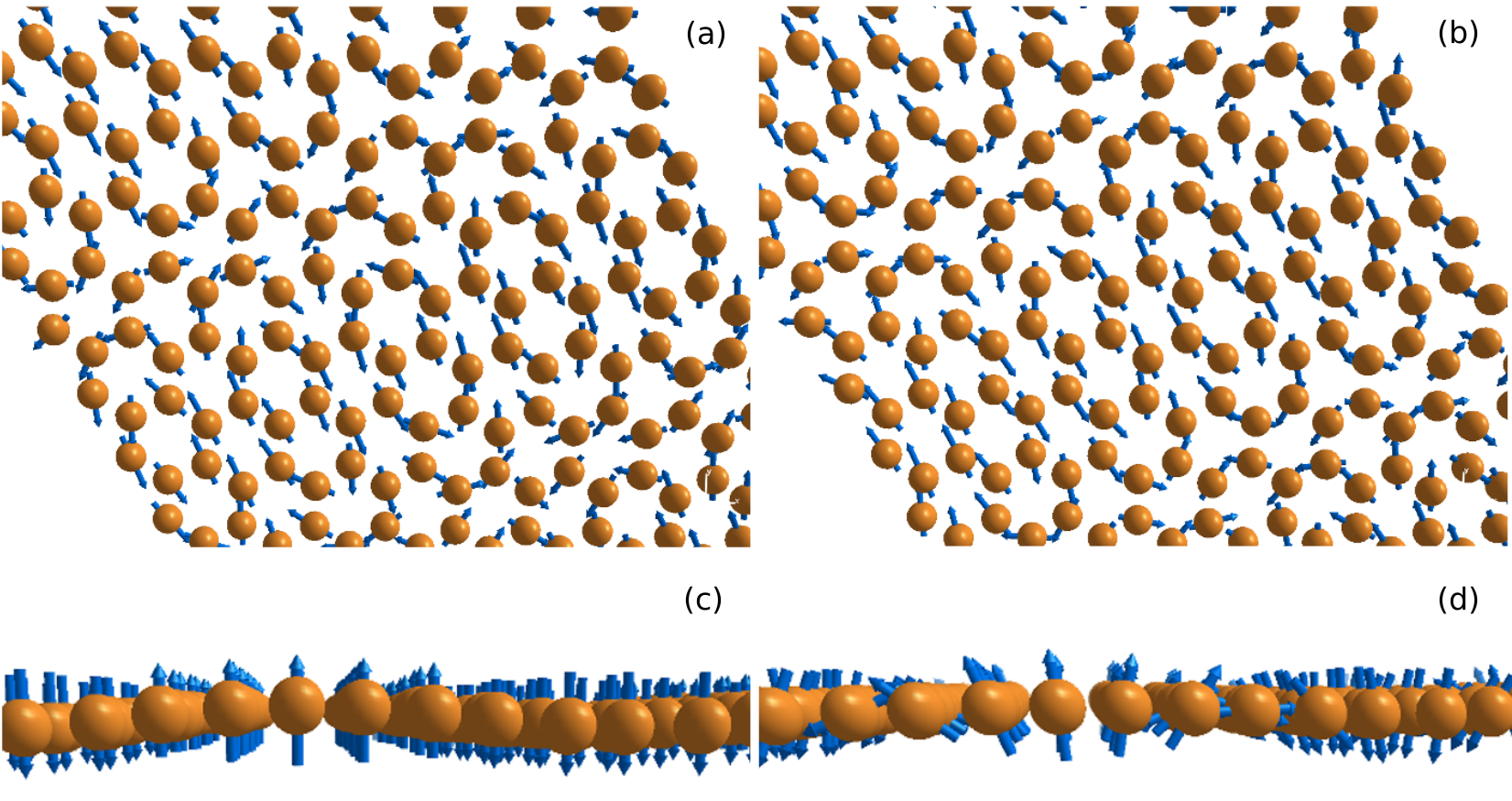}
 \caption{\label{fig-modes} Visualization of MBD eigenmodes for graphene on Ag(111). Orange spheres represent QHOs at the positions of the atoms, arrows depict the polarization direction. Shown are a lateral polarization mode within the graphene sheet for the isolated (a) and adsorbed sheet (b) and a vertical polarization mode, again, for the isolated (c) and adsorbed sheet (d). Contributions from the underlying metal surface are omitted.}
\end{figure*}

The importance of many-body contributions is evident from the collective eigenvectors of the QHO Hamiltonian that represent the eigenstates of the long-range correlation problem in the basis of atomic positions. Although these eigenstates merely constitute the canonical basis of the QHO model Hamiltonian and do not correspond to any  actual physically observable quantities, their change upon adsorption can help to qualitatively analyze the collective mutual polarization and depolarization between different domains of the system. Figure \ref{fig-modes} shows  two such representative MBD eigenmodes (top and bottom) for an isolated graphene sheet and graphene adsorbed on Ag(111) (left and right). The blue vectors indicate the direction and magnitude of polarization on each localized QHO (depicted as orange spheres). The first eigenmode (a and b in Fig. \ref{fig-modes}) describes lateral polarization within the graphene sheet and is almost unaffected upon adsorption on the metal surface. The second mode (c and d in Fig. \ref{fig-modes}) represents polarization orthogonal to the graphene plane and is strongly modified due to adsorption. In fact, this mode is fully delocalized over adsorbate and substrate (not shown here) and describes the collective polarization between the subsystems. These visually apparent changes can also be seen in the energy of the eigenstates. Whereas the energy of the fist eigenmode only decreases by about 1.7~meV upon adsorption, the second mode contributes about 32~meV to the dispersion energy defined by the sum of eigenenergies. Due to the energy shifts of the MBD modes we can in fact pinpoint which polarization modes yield the most important contributions to the dispersion energy. 

\begin{figure*}
 \centering\includegraphics{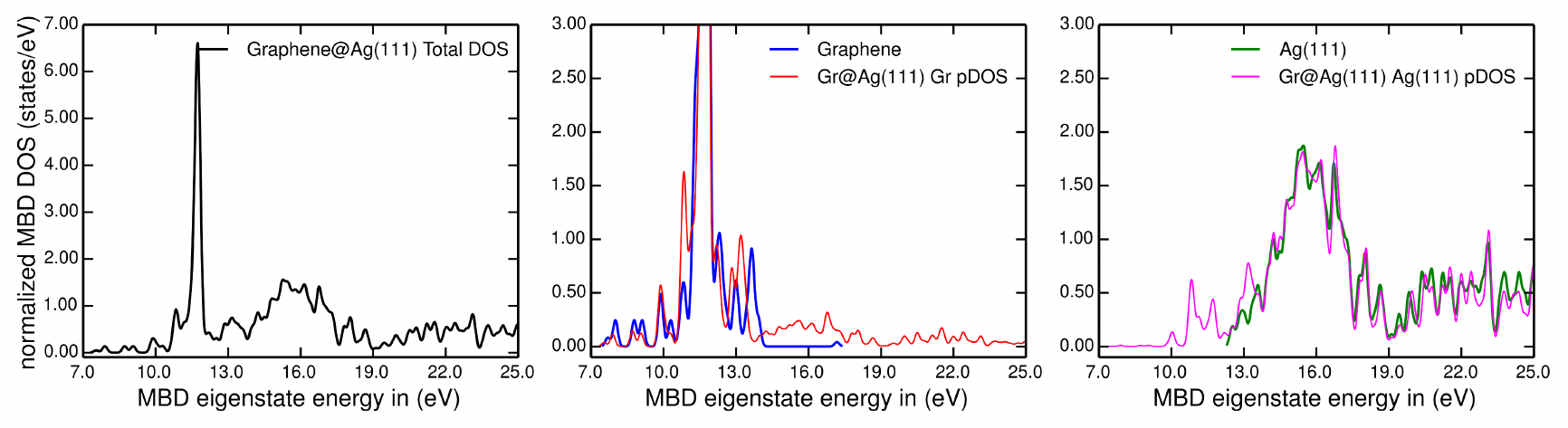}
 \caption{\label{fig-modes-dos} Quasiparticle Density-of-State (DOS) of MBD eigenstates for graphene adsorbed on Ag(111) (left), isolated graphene and the projected DOS of graphene (center, red and blue curves), and a clean Ag(111) surface and the projected Ag(111) DOS (right, green and magenta curves) contributions.}
\end{figure*}

It is possible to visualize the change in MBD modes due to adsorption analogous to a density-of-states (DOS) diagram (see Fig. \ref{fig-modes-dos}). In the MBD-DOS of graphene adsorbed to Ag(111), the first leading peak almost solely stems from polarization modes of the graphene sheet as apparent from the comparison to the DOS of isolated graphene (blue curve in Fig. \ref{fig-modes-dos}), whereas the second large peak almost solely stems from eigenenergies of the substrate. The actual modifications in the graphene DOS due to adsorption can be seen by projecting the isolated graphene states from the spectrum of graphene on Ag(111) (red curve). Whereas the overall DOS of graphene appears basically unchanged, a number of new modes appears at higher energies around 16~eV due to hybridization with modes of the substrate. At the same time polarization modes previously localized to the substrate are shifted towards lower energies (magenta curve) by hybdridization with graphene (at $\sim$ 11.5~eV in Fig. \ref{fig-modes-dos}). While these changes in eigenmodes are instructive to study, it should be noted that they correspond to fluctuations of a model system composed of coupled QHOs. Therefore, unfortunately comparison of the MBD modes to experimental observables is not possible. Only integrated quantities in the MBD model, such as frequency-dependent polarizability or the binding energy, can be directly compared to experiment.

\begin{table}
\caption{\label{tab-graphene} Adsorption energy and vertical height of a graphene sheet adsorbed to Ag(111). Energy is given in meV per carbon atom, the distance is measured as average distance between carbon atoms and the first substrate layer.}
 \begin{tabular}{ccc} \hline
Graphene/Ag(111)	&  -$E_{\mathrm{ads}}$/C atom [meV] & $d$ [\AA] \\ \hline

  PBE+vdW$^{\mathrm{surf}}$	&  72   & 3.05 \\ 
  PBE+MBD			&  45  &  3.23  \\ \hline  
  LDA \cite{Olsen2013}	&  30   &  3.22 \\
  RPBE \cite{Olsen2013}	&   1   &  5.57 \\
  (cRPA+EXX)@PBE \cite{Olsen2013}	&   78   &  3.31 \\ 
  vdW-DF2(C09x) \cite{Loncaric2014}	&   58   &  3.27 \\
  optB86b-vdW-DF \cite{Loncaric2014}	&   67   &  3.34 \\  
  vdW-DF \cite{Vanin2010}	&   33   &  3.55 \\    
  rVV10 \cite{Loncaric2014}	&   68   &  3.48 \\   
  \hline
 \end{tabular} 
\end{table}

The MBD framework enables us to capture the qualitative physics of many-body dispersion and interpret it with concepts familiar to electronic structure theory. On the other hand the quantitative performance of PBE+MBD, in the case of graphene on Ag(111), can only be evaluated in comparison to experimental data. However, due to a lack of such data for graphene on Ag(111), comparison to other simulation approaches can help to put binding energies and adsorption geometries into perspective. While the effective pairwise PBE+vdW$^{\mathrm{surf}}$ approach already improves on the notorious underbinding of pure semi-local functionals, the resulting adsorption energy and height are still above and below what is found when applying non-local van der Waals functionals~\cite{Loncaric2014,Vanin2010,Hamada2010} or RPA-based correlation methods~\cite{Olsen2013} (see Table {\ref{tab-graphene}}). Introducing many-body effects via MBD puts the resulting adsorption height and adsorption energy in the same range as for example optimized van der Waals functionals as proposed by Klimes and Michaelides (optB86b-vdW-DF) \cite{Klimes2011, Loncaric2014}. Results obtained from exact exchange and correlation based on RPA (cRPA+EXX) \cite{Olsen2013} yield an adsorption height similar to what is found with PBE+MBD, but at a surprisingly high adsorption energy per atom, considering that cRPA is typically considered to underestimate binding energies of small intermolecular complexes \cite{Zhu2010}. In fact, the 78~meV/ C atom adsorption energy reported by Olsen \emph{et al.} might not be fully converged with respect to the number of substrate layers, basis-set truncation, and k-point sampling, considering the slow convergence of RPA correlation energies with respect to these parameters \cite{Olsen2013}. Additionally, the neglect of the $q=0$ contribution to the correlation energy might yield an unexpected bias towards higher interaction energies. The overall remaining spread in the adsorption energy and height for graphene/Ag(111) across methods still poses a challenge to be further investigated, for which experimental reference values would be welcome.

\section{Conclusions}
\label{conclusions}

Accurately treating electronic correlations in realistic systems without resorting to highly demanding correlation methods, such as the 'chemistry gold standard' CCSD(T) \cite{Raghavachari1989} or ACFDT-based correlation techniques, can be exceedingly challenging. This is even more so the case for condensed-matter systems, where periodicity, simultaneous existence of localized and delocalized states, and the shear system size, limit the applicability of quantum-chemical approaches. The here discussed DFT+MBD method yields an explicit account of long-range many-body dispersion effects, including instantaneous polarization effects, collective response or screening, which is not only necessary to yield a quantitative account of adsorption energies, but often also is necessary to yield a qualitatively correct description of adsorbate-substrate structure and binding. In this work, we analyzed the importance of many-body dispersion effects for the correct description of adsorption on metal surfaces for the three test cases of Xe, PTCDA, and graphene adsorbed on Ag(111). We furthermore evaluated the accuracy and performance of DFT+MBD in capturing these effects compared to other approaches and experiment.

Although many-body contributions did not strongly alter the interaction strength for noble-gas atom adsorption in the case of Xe on Ag(111), the effects were still significant considering the system size and furthermore considering the role of noble-gas atoms in the development and justification of pairwise dispersion-correction approaches. Most interesting is the strong anisotropy of the atomic polarizability tensor upon adsorption that is captured with MBD. In the cases of PTCDA and graphene, inclusion of many-body effects reduces the overbinding of pairwise dispersion by 20 and 27~meV per adsorbate atom, amounting to a reduction of adsorption energy of approximately 29\% and 25\% for both systems. At the same time the adsorption height was increased by about 0.1 and 0.2~\AA, respectively. This shows that a mere account of pairwise additive dispersion leads to significant overbinding and many-body effects must not be neglected. The effect of higher-order many-body dispersion contributions is directly visible in the MBD eigenstates that describe the delocalized polarization between adsorbate and substrate. The subsystem hybridization visible from the eigenmode analysis can function as an interpretational tool to understand the adsorbate-substrate interactions.

Despite the clear success of the DFT+MBD approach, several methodological (\textit{cf.} Ref. \onlinecite{DiStasio2014}) as well as issues specific to metal-surface adsorption remain. One of which is that the polarizability screening and many-body interactions described by the DFT+MBD method do not yet explicitly feed back into the electronic density or affect the potential that enters the Kohn-Sham equations. As a result the DFT level alignment of adsorbate states with respect to the substrate Fermi level will not be directly affected by the MBD correlation contribution and will still be described at the level of the semi-local exchange-correlation treatment. This effect together with the self-interaction error in the exchange functional will still plague short-range interactions such as covalent bonding. However, in the above presented range-separation framework, density functionals beyond the generalized gradient approximation can also be coupled with the MBD scheme, as was already shown above with the application of HSE+MBD for Xe and PTCDA on Ag(111).

As was shown in section \ref{methods} of this manuscript, in treating adsorbates on metal surfaces DFT+MBD can be understood as combining, so to say, the best of both worlds - of DFT and wavefunction approaches. Whereas the semi-local or screened hybrid density functionals correctly treat the delocalized metallic states of the underlying substrate, the missing long-range correlation is accounted for by recasting the long-range correlation problem into an auxiliary system of coupled QHOs. The close ties of the MBD approach to correlation techniques open a path to systematically go beyond the current state-of-the-art. Replacing the dipole-dipole interactions with an attenuated Coulomb potential, the full machinery of quantum chemistry could be employed to solve the coupled QHO Hamiltonian. The currently employed ``minimal basis'' approach of associating every atom with a single QHO could furthermore be extended by an expansion of the electron density around an atom in terms of a linear combination of QHOs. With such a basis set, the coupled QHO problem can be solved using any correlated quantum-chemical method, all of which would still be at a comparably low computational cost when compared to solving the full electronic many-body problem. 

The above results show that the DFT+MBD approach, although solely developed for describing the correlation effects in molecules and finite-band gap materials, improves the description of molecular adsorption on metal surfaces for the here studied systems. The initially localized QHOs do not correctly account for the response of the metal, however, their interaction leads to a delocalized response that significantly improves the description of long-range correlation. As a result the description of the density-density response becomes closer to a metallic response. In addition to a gain in accuracy, including many-body dispersion effects via the MBD technique offers many conceptual insights into adsorbate-substrate bonding, which, overall, makes it a strong contender in the accurate modeling of complex organic--inorganic interfaces.

\begin{acknowledgments}
R.J.M gratefully acknowledges support by J. C. Tully (Yale University) and the U.S.  Department of Energy, Basic Energy Sciences grant DE-FG02-05ER15677. V.G.R. and A.T. thank the European Research Council for financial support (ERC-StG VDW-CMAT).
\end{acknowledgments}


%

\end{document}